\documentclass[twocolumn,prl,aps,superscriptaddress]{revtex4}
\pdfoutput=1
\usepackage{amsmath,amssymb,latexsym,revsymb}
\usepackage[pdftex]{graphicx}
\usepackage{epsfig}

\def\duzomniejsze{<\kern-.7mm<}
\def\duzowieksze{>\kern-.7mm>}

\def\textbf#1{{\bf #1}}
\def\beq{\begin{equation}}
\def\eeq{\end{equation}}
\def\be{\begin{equation}}
\def\ee{\end{equation}}
\def\ben{\begin{eqnarray}}
\def\een{\end{eqnarray}}
\def\beqa{\begin{eqnarray}}
\def\eeqa{\end{eqnarray}}
\def\eea{\end{array}}
\def\bea{\begin{array}}
\newcommand{\bei}{\begin{itemize}}
\newcommand{\eei}{\end{itemize}}
\newcommand{\bee}{\begin{enumerate}}
\newcommand{\eee}{\end{enumerate}}

\newcommand{\tr}{\operatorname{Tr}}

\def\>{\rangle}
\def\<{\langle}
\def\qed{\vrule height 4pt width 3pt depth2pt}

\def\rab{\rho_{AB}}
\newcommand{\ket}[1]{| #1 \rangle}
\newcommand{\bra}[1]{\langle #1 |}

\newtheorem{lemma}{Lemma}

\newtheorem{theorem}[lemma]{Theorem}
\newtheorem{definition}[lemma]{Definition}
\def\psiabe{\ket\psi_{ABE}}
\def\12{{\textstyle \frac{1}{2}}}
\def\s{\,\,\,}
\def\ent{{E}}
\def\minent{{\overline E}}

\def\q{{\overline Q}}
\def\qund{{\underline Q}}
\def\E{{\tilde E}}

\newcounter{protoline}
\newlength{\boxwidth}
\newlength{\bigboxwidth}

\newtheorem{proto_body}{Protocol}
\newenvironment{protocol}[1]{ 
\setcounter{protoline}{0}
\begin{minipage}{\boxwidth}

\begin{proto_body}[ #1 ]
\ \\[0mm] 
\begin{description} }{\end{description}\end{proto_body}\end{minipage}}
\newenvironment{protocol_cont}[0]{ 
\begin{minipage}{\boxwidth}
\addtocounter{proto_body}{-1}

\begin{proto_body}
\ \\ 
\begin{description} }{\end{description}\end{proto_body}\end{minipage}}

\begin{document}

\title{A paradigm for entanglement theory based on quantum communication}

\begin{abstract}
Here it is shown that the squashed entanglement has an operational meaning --
it is the fastest rate at which a quantum state can be sent between two
parties who share arbitrary side-information.
Likewise, the entanglement of formation and entanglement cost is shown to be the fastest rate at which a quantum state can be sent when the parties have access to side-information which is maximally correlated.  A further restriction on the
type of side-information implies that the rate of state transmission is given by the quantum mutual information.  This
suggests a new paradigm for understanding entanglement and other correlations.  Different types of side-information 
correspond to different types of correlations with the squashed entanglement being one
extreme.  The paradigm also allows one to classify states not only in terms of how much quantum communication is needed 
to transfer half of it, but also in terms of how much entanglement is
needed.  Furthermore, there is a dual paradigm: if one distributes the side-information as
maliciously as possible so as to make the sending of the state as
difficult as possible, one finds maximum rates which give
interpretations to known quantities (such as the entanglement of
assistance), as well as new ones.  The infamous additivity questions can also be recast and receive an operational
interpretation in terms of maximally correlated states.
% (such as the {\it puffed entanglement} - which is dual to the squashed entanglement).
%We find that the puffed entanglement is equal to the entanglement of assistance, and is thus also dual to the 
%entanglement cost.
%
%The later quantity provides a way
%to quantify the classical correlations, as it is a complement of the entanglement of purification.

\end{abstract}

\author{Jonathan Oppenheim}
\affiliation{Department of Applied Mathematics and Theoretical Physics, University of Cambridge U.K.}

\maketitle

There are two main paradigms for understanding entanglement between two
parts of a system.  We say that a
system is in an entangled state if measurements on it cannot be simulated by a
local hidden variable model.  Another (and perhaps inequivalent way)
to understand entanglement is to say that we do not know what
entanglement is, but we do know that it is a type of correlation which
cannot increase under local
operations and classical communication (LOCC) \cite{Werner1989}.
Entanglement measures are thus monotones (they must go down under LOCC).

Here, we will show that two of the most prominent entanglement measure
and one more general correlation measure have an operational meaning.  They give
the rate at which one share of a state can be sent to a receiver -- broadly speaking
the more entangled a state is, the harder it is to send it.
Here, the rate is how much quantum communication is needed to send
each state, and which
correlation measure corresponds to this rate is determined by what resource is
given to the sender and receiver.  In particular, the resource that
is given to the sender and receiver is side-information 
-- i.e. we give the sender and receiver
quantum states which contain information about the state which is
being sent.  Different restrictions on the states which contain the side information give
different rates at which quantum states can be sent, and these rates turn out to
correspond to different correlation measures.

Two of the correlation measures
($E_c$ the entanglement cost \cite{BDSW1996,cost} and $I(A:B)$ the quantum mutual information) had other operational
meanings, but the squashed entanglement~\cite{Winter-squashed-ent,Tucci02} $E_{sq}$ had thus far been 
a purely formal quantity as the quantum analog of the intrinsic
mutual information~\cite{maurer99unconditionally}:
\beq
E_{sq}\equiv \inf_\Lambda \12 I(A:B|\Lambda(E))
\eeq
with $\Lambda$ a completely positive trace preserving map,   $\psiabe$
a pure state, $I(A:B|\E)=S(A\E)+S(B\E)-S(AB\E)-S(\E)$ the
conditional mutual information of a tripartite state $\rho_{AB\E}$ and
$S(A)$ the entropy of the reduced state  $\tr_{BE}\rho_{AB\E}$.
Here, we find that the squashed
entanglement is an extreme case: it is the rate at which a share of
the state can be sent when the sender and receiver are given the best
possible side information.  The squashed entanglement is a remarkable entanglement measure 
because of it's additivity, elegance and simple expression.  We now
see that it has a very intuitive and operational meaning as well.

This inspires the introduction of a third paradigm in which to
understand entanglement and other correlations.  We can in general 
consider the rate at which states can be sent to a receiver given different types of
restrictions on the side-information states.   This gives a relationship between sets
of states and correlation measures. Furthermore, there is a dual paradigm.  Instead of
considering the best rate at which information can be sent, one can find
the worst possible rate.  I.e. the side-information is distributed as maliciously as possible
in order to make the information as unhelpful as possible to the sender and receiver.  This
leads us to discover a set of operational quantities, some of which
had already been known to be of significance, and some of which were
previously unknown.  We also are able to recast well known additivity questions~\cite{shor-additivity} concerning
the entanglement of formation and other quantities into an operational question about
the utility of certain types of side information.

Let us imagine that Alice and Bob share many copies of a quantum system in state $\rab$, 
and we ask how many qubits are required for Alice to send her share of the state to a third party, Charlie.
Everything we discuss will be symmetric
under exchanging the roles of Alice and Bob.
Here, Bob will be completely passive, his only role being to hold his
part of the state $\rho_B$.  
However, for Alice to successfully send her share of the state, one requires that her protocol work
for all possible pure states she may hold.  Here, Alice knows the statistics of the state (i.e. she knows $\rab$),
but not which particular pure state she holds.

We are interested in maximising the rate at which the state can be sent, and so 
we allow Alice and Charlie to have 
as much help as possible in the form of ancillas.  I.e we give them as much 
additional information as is allowed by the laws of quantum mechanics.  In the classical world, this would
be a lot, since there is nothing to prevent Charlie from knowing exactly what Alice has and therefore nothing would
need to be sent from Alice to Charlie.  However, quantumly,
the amount of information Charlie can have is limited by the fact that Alice's state is entangled with Bob's.
In particular, if we imagine the purification of $\rab$, i.e. a pure state $\psiabe$ such that 
$\tr_E \ket{\psi_{ABE}}\bra{{}_{ABE}\psi} = \rab$, then the most information that Charlie could possible have would be to possess the
share $\rho_E$.  This might not be much information -- for example, if Alice and Bob share a pure state, than Charlie's 
state must be completely uncorrelated from Alice's.

Now if Charlie has $\rho_E$ then the amount of quantum communication that is required when no classical
communication is allowed is given by the mutual information $I(A:B)/2$
\cite{fqsw}, with the protocol for
doing this given by a coherent version of merging~\cite{how-merge}.
Note that the usual convention
is for Bob to be a reference system $R$ and for the receiver to be called Bob, but here this conflicts with the 
convention of quantifying the entanglement between two parties called Alice and Bob.

Now the rate might not be maximised by giving Charlie all the side information -- it may be beneficial
for the sender Alice to have some as well.  One may therefor want to split $\rho_E$ into two shares, some of which
is given to the receiver Charlie (call this $\rho_C$), and some of which is given
to Alice (call this $\rho_{A'}$).  What is the best rate we can
achieve if we optimally give side information to the sender and receiver?

It is simple to show that the answer is the squashed entanglement.  
Given a noiseless quantum channel and arbitrary shared entanglement
between a sender and receiver let $Q_{A'C}$ be the rate of quantum communication that is 
required to send $\rho_A$ to a receiver who holds state $\rho_C$
with $\rho_{A'}$ being held by the sender. Then
\begin{theorem}Given a state $\rab$ and arbitrary side-information
  $\rho_{A'C}$, $\inf Q_{A'C}=E_{sq}(A:B)$ where the infimum is taken
  over states $\rho_{A'C}$ and $E_{sq}$ is the squashed entanglement.  
\end{theorem}

\begin{figure}
\centering
  \includegraphics[width=8cm]{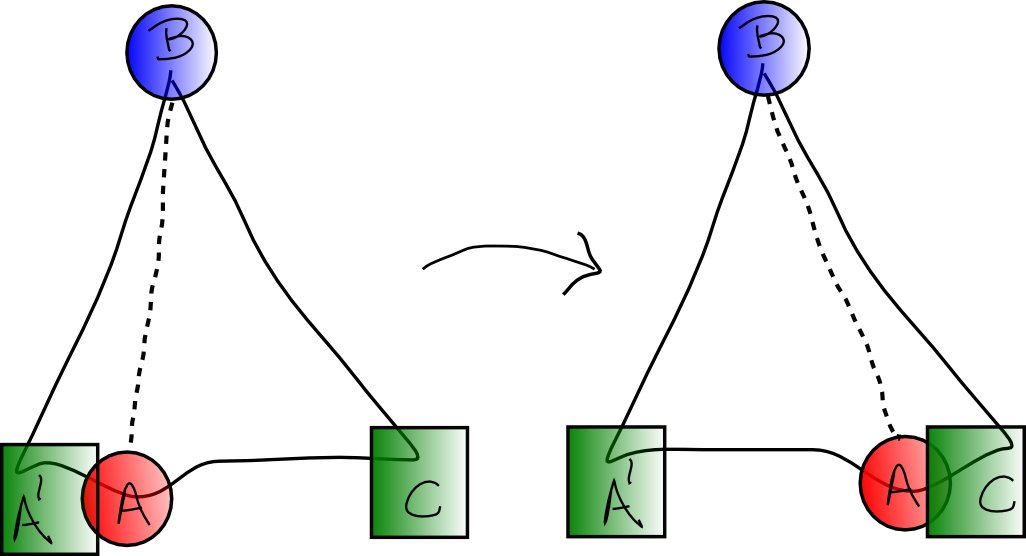}
   \caption{Redistribution: $\rho_A$ is sent to $C$ retaining it's correlations with $\rho_{A'BC}$. 
By optimising over how we split $\rho_{A'C}$ we isolate the entanglement
shared between $A$ and $B$ (dashed line)}
  \label{fig:redis-sketch}
\end{figure}

The problem
of finding the rate for state merging when both the sender and
receiver have side information was found in \cite{devetakyard-redistribution} and was called {\it
state redistribution}. 
Namely Alice only 
wants to send part of her state to Bob who shares part of her state.  The situation is depicted in figure 
\ref{fig:redis-sketch}.

The minimum amount of quantum communication required is given by the optimal cost pair
\begin{eqnarray}
Q_{A'C} & = & \12 I(A:B|C) \label{redistpairQ}\\
\ent &=& \12 I(A:A') - \12 I(A:C). \label{redistpairE}
\end{eqnarray}
with $\ent$ being the amount of entanglement that is required. 
%and $I(A:B|C)=S(AC)+S(BC)-S(ABC)-S(C)$ the quantum conditional mutual
%information. 
A simple explanation for this rate in terms of state
merging is given in \cite{jredistribution}.

Now observe that because $\rho_C$ is a splitting of $\rho_E$ into two parts, $\rho_C$ and $\rho_{A'}$, we can treat
the system $A'$ as an environment for a channel $\Lambda$ (i.e. completely trace preserving positive map) which acts on 
$\rho_E$ and produces the output $\rho_{C}$.  Now, because we give Alice and Charlie all 
side information they require, this includes shared entanglement, and therefore the rate of Equation (\ref{redistpairQ})
can be obtained.  If we now minimise the amount of quantum communication required as
a function of all possible splittings of the ancillas, we have that
this minimal communication $\q_{all}\equiv \inf Q_{A'C}$ is given by
\beqa
\q_{all}&=& \inf_{\rho_{A'C}} \12 I(A:B|C) \nonumber\\
&=& \inf_\Lambda \12 I(A:B|\Lambda(E)) \nonumber\\
&=& E_{sq}(A:B) \s .
\eeqa
Here, the last line follows from the definition of the squashed entanglement.
%$E_{sq}$ is defined as the minimum conditional mutual information over
%possible extensions of the state $\rho_{AB}$.  An extension is any $\rho_{\tilde E}$ such that 
%$\tr_{\tilde E} \rho_{AB{\tilde E}}=\rab$, and thus
%can be written as a channel acting on the purification $\rho_E$.
\qed

This gives an operational meaning to the squashed entanglement.  The amount of pure state entanglement, $\ent$, that
is gained given the ancilla splitting which minimises $Q_{A'C}$ gives
us a second parameter $\minent$ which would appear to characterise
the state.  Using the purity of the total state, and the fact that $S(F)=S(G)$ for a total pure state $\psi_{FG}$, we
can convert Equation (\ref{redistpairE}) into
\beq
\ent = S(A|C)-Q
\eeq
with $S(A|C)$ the conditional entropy $S(AC)-S(C)$ and $\ent$ being the entanglement that is consumed during the protocol
(negative values indicate that entanglement is gained).  Not surprisingly, the total amount of shared states i.e.
$Q_{A'C}+E$ gives the amount of entanglement required to merge \cite{how-merge} $\rho_A$ with the ancilla $\rho_C$.
Even a state with zero squashed entanglement (thus 
$\q_{all}=0$) may consume $S(A|C)$ bits of pure state entanglement to share (or may result in a gain of ebits).  
Minimising this quantity as well
\beq
\minent_{all}=\inf_C S(A|C)-\q_{all}
\eeq
where the infimum is taken over all $C$ which minimise $\q_{all}$
gives a way to differentiate even 
between states with the same squashed entanglement.  We will see that this quantity is zero for separable
states, although there may be some other states for which the squashed entanglement is zero, but for which $\minent_{all}$
is non-zero.  We will also find that this quantity is zero for any state where the squashed entanglement is equal to the entanglement
cost.

It is worthwhile trying to understand why entanglement between Alice
and Bob should effect the rate at which Alice's share can be sent to a receiver. 
We noted previously that monogamy of entanglement plays a crucial role here.  If Alice's system is entangled with Bob, 
then it will be more difficult to find side-information on the ancillas $A'$ and $C$ which can be correlated with
$\rho_A$ in such a way as to make the task of sending $\rho_A$ easier.  There is another reason why the procedure
above quantifies correlation.  Sending $\rho_A$ is deemed successful if Alice's share of the state is at the
receiver, and the fidelity of the total pure state $\psiabe$ is kept
during the protocol.  This is equivalent to the protocol working for
any pure state decomposition of $\rab$ (since measurements on $\rho_E$
will induce a pure state ensemble on $\rab$.  To preserve the overall
pure state, Alice must
thus send her state while keeping it correlated with Bob's.  The more correlation there is with Bob, the more
she has to send.  We can think of Bob as being the referee who checks to make sure that Alice has actually sent
her state to Charlie (he would check this by bringing his share
together with the other shares and measuring the total state).  
The more correlation Bob has with Alice, the more she must send.  
%A classical
%example of this is if Alice shares perfect correlations with Bob, and
%must send her part to Charlie.  Bob can test that Alice has sent her
%share by checking that his system is now correlated with what Charlie
%has.     

Given this intuition, we can see if it leads to a more general paradigm -- quantifying a state's correlation with a 
another system 
by how much has to be sent to transfer the state and maintain the correlation. If we allow arbitrary side-information
at the sender and receiver, then the best rate of sending is given by the squashed entanglement $E_{sq}$.  We can 
generalise this to consider more restricted types of side-information.  $E_{sq}$
thus represents one possible extreme.  We will see that another correlation measure, the mutual information, 
is the other extreme.  In between, we have other measures such as
entanglement cost.  We will also see that there is a dual paradigm.  Table 1 gives a summary 
of the results.

\begin{table*}[ht]
\begin{tabular}{|l|l|l|}
\hline
side-information & infimum rate & supremum rate \\ \hline\hline
no restriction 
& $E_{sq}$ (squashed entanglement) 
& $E_{pu}$ (puffed entanglement)* \\ \hline
maximally correlated states (MCS) & $E_c$ (entanglement cost) &
$E_{ass}$ (entanglement of assistance) \\ \hline
all of it at the receiver or sender & $I(A:B)/2$ (mutual information)
&$I(A:B)/2$ (mutual information) \\
\hline
\end{tabular}
\caption{Summary of relationships between the restrictions on
side-information and the rate at which half the state can be sent to
the receiver.  Note that $E_{pu}=E_{ass}=\min\{S(A),S(B)\}$.  *$E_{pu}$ is
the supremum rate if pure entanglement is an allowable resource.}
\end{table*}

We consider the quantity $\q^{O}_{\cal S}$ to be the minimum qubit rate for sending $\rho_A$ to a receiver, using side information 
at the encoder and decoder $\rho_{A'C}$ chosen from the set ${\cal S}$, and the allowed class of operations
being $O$.  In this paper we will mostly consider $O$ to be local operations and the sending of qubits which is quantified,
so we will thus drop the superscript.  However, one might also want to consider $O$ which include free entanglement or 
classical communication.
\begin{definition}
%the  $\q_{\cal S}$ is 
\beq
\q_{\cal S}\equiv \inf_{\rho_{A'C}\in {\cal S}} Q_{A'C}(A:B)
\eeq
\end{definition}
However, in general one can consider any class of operations, and one can even count some other resource (e.g. private
communication) rather than sent qubits.  We have so far considered the case when
the set ${\cal S}$ is unlimited and contains all possible states, including 
possible extensions of $\rho_{AB}$.

Let us now consider the case where the set ${\cal S}$ is restricted to the set of maximally correlated states (MCS).  
I.e.
states of the form 
\beq
\rho_{A'C}=\sum_{ij}\sigma_{ij} \, {} \ket{ii}\bra{jj} \s .
\label{eq:mcs}
\eeq
With such a restriction on the side information $\rho_{A'C}$, 
we find that the maximum rate for sending $\rho_A$ is given by the entanglement cost $E_c$. 
\begin{theorem}Given a state $\rab$, $\q_{\cal S}=E_{c}(A:B)$ when ${\cal S}$ is the set of 
  MCS.   
\label{thm:ec}
\end{theorem}
$E_c$ 
is defined as the amount of pure state entanglement which is required to create a copy of the state in the limit
that many copies of the state are created, and it is given by
\beq
E_c=\lim_{n\rightarrow\infty}\inf \sum_i p_i \frac{S(\rho_A^i)}{n} 
\eeq
where the infimum is taken over decompositions
\beq
\rab^{\otimes n}=\sum_i p_i \ket{\psi^i_{AB}}\bra{\psi^i_{AB}} 
\label{eq:ecdecomp}
\eeq
(the entanglement of formation is defined as the entanglement cost, but for single copies of $\rab$, i.e. $n=1$).

To prove Theorem \ref{thm:ec}, we note that for any decomposition (not just the one which minimises the entanglement
of formation), we can write the purification as
\beq
\ket{\psi}_{ABA'C}^{\otimes n}=\sum \sqrt{p_i} \ket{\psi^i}_{AB}\ket{ii}_{A'C} \s .
\label{eq:mcspureification}
\eeq
One can verify that the reduction of this state on $A'C$ is a maximally correlated state with 
$\sigma_{ij}=\sqrt{p_i}\sqrt{p_j}\bra{\psi^i}\psi^j\rangle$.  Furthermore for MCS we find, 
\beq
I(A:B|C)/2=\sum_i p_i S(\rho_A^i) \s .
\eeq
This quantity is equal to the rate for sending Alice's share to Bob when entanglement is allowed 
as a resource, however, because pure state entanglement can be in the form of a MCS, it gives the rate in this case
as well.
Taking the infimum of this expression over decompositions of Eq. (\ref{eq:mcspureification}) 
gives $\inf I(A:B|C)/2=n E_c$.  Since all maximally correlated states can be written as the reduction
of a purification of the form in Eq. (\ref{eq:mcspureification}) we have that the infimum over 
decompositions (\ref{eq:ecdecomp}) is equivalent to an infimum over maximally correlated states.  \qed
We thus have another
interpretation of the entanglement cost: it is the
fastest rate at which $\rho_A$ can be sent to a receiver using side-information composed of maximally
correlated states.

Note that the entanglement of formation 
has a similar interpretation if we restrict ourselves to side-information which is an extension of 
each single copy.  While this doesn't matter for the squashed entanglement which is additive, we do
not know whether it matters for the entanglement of formation. Here, the additivity question is whether using 
maximally correlated states which are the purification of two states is no better 
for state transfer, than maximally correlated states which are the purification of each single state.  We thus
have another additivity question related to the others -- one that is operational.

Also, the entanglement cost $\ent$ when the side-information is restricted to MCS is always zero.  This can be 
seen from expression (\ref{redistpairE}) and the fact that MCS are symmetric.  It is for these reason that
$\ent$ is zero whenever the squashed entanglement is equal to the entanglement cost.

The most simple example of the paradigm is when the set $\cal S$ are states
such that all the side information is at the sender or the receiver.  In
such case, it is trivial to see that the communication rate is the mutual
information $I(A:B)/2$ which is a measure of total correlations (both
classical and quantum).  Note that if we bend the paradigm slightly, and allow
for the restriction on ${\cal S}$ to allow for no distribution of side-information
then the best rate that Alice can achieve is through simple compression, giving a rate 
of $S(A)$.

Thus far, we have been distributing the side information so as to
minimise 
the amount of communication required.
There is a dual paradigm -- instead we
arrange the side-information in the worst possible manner.
I.e.
\begin{definition}
the dual to $\q_{\cal S}$ is 
\beq
\qund_{\cal S}\equiv \sup_{\rho_{A'C}\in {\cal S}} Q_{A'C}(A:B)
\eeq
\end{definition}
Here, the distributor of the side information is forced to distribute
all the side information, so the only choice they have is how to arrange it between
Alice and Charlie.
When we do this, we will find that the rates for state transfer are given by dual 
correlation measures.  For example, the
entanglement of assistance~\cite{entass} is the maximum amount of pure state
entanglement that can be created between Alice and Bob given a
measurement on the purification of their state.   Here, we find another
interpretation -- it is the amount of information which must be sent
if the side-information $\rho_{A'C}$ is maliciously chosen from the set
of maximally correlated states.
\begin{theorem}
%The amount of quantum communication required to send
%  $\rho_A$ to the receiver $C$ is given by the entanglement of
%  assistance when this rate is maximised over side information 
%$\rho_A'C$ chosen from  the set of MCS.
Given a state $\rab$, $\qund_{\cal S}=E_{ass}(A:B)$ when ${\cal S}$ is the set of 
  MCS.   
\label{thm:eass}
\end{theorem}

The proof is similar to that of the
entanglement cost except that one needs to use the fact that $\ent=0$ for MCS since a malicious distributor
of the side information is unlikely to give the parties any entanglement.
% (the
%entanglement of assistance is super-additive~\cite{eofa}[quant-ph/9803033] therefore it doesn't
%matter whether one considers side information on the single copy level
%or on multiple copies).  What is more, we know which MCS is the worst one.
The entanglement of assistance is known to be equal to
$\min\{S(A),S(B)\}$  \cite{svw2005,how-merge}.  An
optimal protocol~\cite{how-merge} for creating entanglement between Alice and Bob is to
perform a random measurement of dimension $S(E|\min{A,B})$ on the
purification.  This
implies that the worst possible maximally correlated state is one
where there are just under $S(E|\min\{A,B\})$ basis elements $\ket{i}_C$, and they
are chosen at random over the typical space of $\rho_E^\otimes n$. 

It is tempting to look at maximising the amount of communication
required over all possible side-information states.   This quantity
would be a dual to the squashed entanglement but with the infimum changed
to a supremum.  
\begin{definition} The {\it puffed entanglement}\footnote{I apologise
    for this name, but what exactly would you expect the dual of the
  squashed entanglement to be called?} of the state
  $\rab$ is $E_{pu}=\sup \frac{1}{2}I(A:B|{\tilde E})$, with the supremum
taken over all extensions ${\tilde E}$ of $\rab$.
\end{definition} 
One can show that it is not an entanglement measure, but it's operational meaning is
that it is the rate required to send the state $\rho_A$ if the
distributor of the side information is as malicious as possible, and
distributes $\rho_{A'C}$ in as unhelpful  a way possible.  However, here one does
have to allow the parties free entanglement.  

Remarkably, the puffed entanglement is equal to the entanglement of
assistance.  
\begin{theorem}
%The amount of quantum communication required to send
%  $\rho_A$ to the receiver $C$ is given by the entanglement of
%  assistance when this rate is maximised over side information 
%$\rho_A'C$ chosen from  the set of MCS.
Given a state $\rab$, $\qund_{\cal S}=E_{pu}(A:B)=\min\{S(A),S(B)\}$ 
when ${\cal S}$ is the set of all possible states.   
\label{thm:puff}
\end{theorem}
The second equality comes because
standard entropic inequalities imply that  
$I(A:B|C)/2\leq \min\{S(A),S(B)\}$, while a protocol exists (using an
MCS) to get equality.  Thus the same side information
which maximises the entanglement of assistance also maximises the
puffed entanglement.  It is perhaps surprising that getting to choose
the side information over
all possible states but with free entanglement for the parties, is equivalent to 
choosing the side information
from maximally correlated states. One can also show that another way of selecting 
the worst possible side-information is to perform a random unitary on
$\rho_E^{\otimes n}$ and then choose $\rho_A'$ of size just over $\min
\{nI(E:A)/2,n I(E:B)/2\}$ qubits.  This means that $\rho_C$ will be
decoupled from both $\rho_A$ and $\rho_B$.  It is an open question what the
supremum sending rate is if entanglement is not a free resource.

Both the entanglement of assistance, and puffed entanglement are
super-additive because they involve an 
infimum over possible extensions.  However, neither are additive i.e.
one can have 
\beq
E_{pu}(\rab^1\otimes\rab^2)> E_{pu}(\rab^1)+E_{pu}(\rab^2)
\eeq
a situation which can occur when, for example, $S(A_1)<S(B_1)$ and
$S(B_2)>S(B_2)$.

For the case of side-information all at the sender or receiver, the rate is half the mutual information 
regardless of whether we want to minimise or maximise the
communication rate, so in this paradigm, the mutual information is self-dual.
%for this restriction on $\cal S$, the puffed communication rate is equal
%to the optimised communication rate.

It would be interesting to look at the correlation measures and rates
under other restrictions
of the set ${\cal S}$.  For example, one could consider side-information which is separable, or whose
states have positive partial transpose, or which are only classically correlated.  Likewise, one might
also consider other types of operations.  As one potential example,
there appears to be a strong link between the intrinsic mutual
information~\cite{maurer99unconditionally} and the maximum rate of sending a private distribution
using only a private channel~\cite{jredistribution}.  Here, the intrinsic
information is a measure of the maximum rate for sending 
classical distributions if one stays within a certain class of
distributions~\cite{oc}.  
%The extent to which this property
%holds for general distributions is unclear.

It would be interesting to know what the conditions are on the set ${\cal S}$ such that $\q_{\cal S}$
is an entanglement measure (or correlation measure).  Some small step towards this is given by the theorem
\begin{theorem}
If ${\cal S}$ includes pure entanglement and 
for all $\sigma_0,\sigma_1 \in {\cal S}$ and all purifications $\ket{\psi_0}$,$\ket{\psi_1}$
$\tr_X(\ket{\psi_0}\ket{00} + \ket{\psi_1}\ket{11})/\sqrt{2}\in{\cal S}$
then $\q_{\cal S}$ is an entanglement monotone.  Here the trace is taken over the Hilbert space which is
complement to $\sigma_0,\sigma_1$ (i.e. the purification).
\end{theorem}
The above theorem is enough to imply that $\q_{\cal S}$ is convex.  One then only needs to show that
$\q_{\cal S}$ cannot increase on average after application of a local CPT map~\cite{Vidal-mon2000}.  The adaptation
of the proof of Theorem 11.15 of \cite{Nielsen-Chuang} 
related in \cite{Winter-squashed-ent} is sufficient for our purposes.

In general, we have found a relationship between various sets of states
and correlation measures.  Here, the squashed entanglement arises as a
more quantum version of the entanglement cost.   While traditionally,
these correlation measures arise as a monotone under some class of operations, here everything is
determined in terms of the set of states.  Remarkably, the notion of classical communication or classicality
does not enter into the discussion, yet it gives arise to at
least two entanglement measures.

There is a sense in which the paradigm is reversible: rates are the same regardless of who gets what part of the
side-information state, i.e. if we 
swap $A'$ with $C$.  This means that the same amount of communication is needed to send the state from Alice to Charlie as from Charlie
to Alice, and we can send the state back and forth using and recovering the pure state entanglement.

It would also be interesting to explore the significance of
the amount of entanglement which is consumed at the maximal (or dual) sending rate.
It may even shed some light on the relationship between the paradigm
considered here, and that of LOCC.  In this case, a key question is
whether a state that doesn't show its entanglement here (in the sense
that it requires no quantum communication to send half of it to a
receiver), is also unentangled in the LOCC case.  We know that if
$E_c=0$ then $E_{sq}=0$, but the converse could well be false.  It would also be interesting to
know what the dual to the squashed entanglement is in the case where pure entanglement is not an allowable resource.

\begin{acknowledgments}
This work is supported by 
EU grants QAP and by the Royal Society.  I am grateful to Micha{\l} Horodecki for inspirational discussions.
\end{acknowledgments}
\bibliographystyle{apsrev}
%\bibliography{c:/Prace/Referencje/refmich,c:/Prace/Referencje/refjono}
\bibliography{../refmich,../refjono,../refjono2}

%@article{ maurer99unconditionally,
%    author = "Ueli Maurer and Stefan Wolf",
%    title = "Unconditionally Secure Key Agreement and the Intrinsic Conditional Information",
%    journal = "{IEEE} Transactions on Information Theory",
%    volume = "45",
%    number = "2",
%    pages = "499--514",
%    year = "1999",
%    url = "citeseer.ist.psu.edu/maurer99unconditionally.html" }

\end{document}